\documentclass[10pt, conference]{IEEEtran}
\IEEEoverridecommandlockouts

\usepackage{xparse}
\usepackage{xspace}
\usepackage{comment}
\usepackage{xcolor, soul}
\usepackage{xcolor,colortbl}

\usepackage{enumitem, kantlipsum}
\usepackage{filecontents,lipsum}
\usepackage[noadjust]{cite}
\usepackage{graphicx}
\usepackage[hidelinks]{hyperref}
\usepackage{MnSymbol}

\newcommand{\dataset}{{\fontsize{1.2em}{60}\selectfont S}OSum\xspace}

\newcommand{\postnum}{2,278}

\usepackage{xcolor}
\usepackage{caption}
\usepackage{tabularray}
\UseTblrLibrary{booktabs}

\usepackage{svg}
\usepackage{booktabs}
\usepackage{array}
\newcommand{\PreserveBackslash}[1]{\let\temp=\\#1\let\\=\temp}
\newcolumntype{C}[1]{>{\PreserveBackslash\centering}p{#1}}
\newcolumntype{R}[1]{>{\PreserveBackslash\raggedleft}p{#1}}
\newcolumntype{L}[1]{>{\PreserveBackslash\raggedright}p{#1}}

\definecolor{Gray}{gray}{0.9}
\definecolor{light_green}{HTML}{c4edc9}
\definecolor{light_red}{HTML}{fddad9}

\DeclareRobustCommand{\hlred}[1]{{\sethlcolor{light_red}\hl{#1}}}
\DeclareRobustCommand{\hlgreen}[1]{{\sethlcolor{light_green}\hl{#1}}}

\newcommand{\howtonum}{301}
\newcommand{\whatisnum}{305}
\newcommand{\debugnum}{265}

\newcolumntype{N}[1]{>{\centering\arraybackslash}m{#1cm}}

\usepackage{color, colortbl}

\newcommand{\name}{\textsc{Assort}}
\newcommand{\liu}{\textit{BertSum}}
\newcommand{\uname}{\textsc{Assort}$_{IS}$}
\newcommand{\sname}{\textsc{Assort}$_{S}$}

\AtBeginDocument{%
  \providecommand\BibTeX{{%
    \normalfont B\kern-0.5em{\scshape i\kern-0.25em b}\kern-0.8em\TeX}}}

\begin{document}



\title{Automated Summarization of Stack Overflow Posts\\}
\author{\IEEEauthorblockN{Bonan Kou}
\IEEEauthorblockA{
\textit{Purdue University}\\
West Lafayette, USA \\
koub@purdue.edu}
\and
\IEEEauthorblockN{Muhao Chen}
\IEEEauthorblockA{
\textit{University of Southern California}\\
Los Angeles, USA \\
muhaoche@usc.edu}
\and
\IEEEauthorblockN{Tianyi Zhang}
\IEEEauthorblockA{
\textit{Purdue University}\\
West Lafayette, USA \\
tianyi@purdue.edu}
}

\maketitle


\begin{abstract}
Software developers often resort to Stack Overflow (SO) to fill their programming needs. Given the abundance of relevant posts, navigating them and comparing different solutions is tedious and time-consuming. Recent work has proposed to automatically summarize SO posts to concise text to facilitate the navigation of SO posts. However, these techniques rely only on information retrieval methods or heuristics for text summarization, which is insufficient to handle the ambiguity and sophistication of natural language.

This paper presents a deep learning based framework called {\name} for SO post summarization. {\name} includes two complementary learning methods, {\sname} and {\uname}, to address the lack of labeled training data for SO post summarization. {\sname} is designed to directly train a novel ensemble learning model with BERT embeddings and domain-specific features to account for the unique characteristics of SO posts. By contrast, {\uname} is designed to reuse pre-trained models while addressing the domain shift challenge when no training data is present (i.e., zero-shot learning). Both {\sname} and {\uname} outperform six existing techniques by at least 13\% and 7\% respectively in terms of the F1 score. Furthermore, a human study shows that participants significantly preferred summaries generated by {\sname} and {\uname} over the best baseline, while
the preference difference between {\sname} and {\uname} was small. 
\end{abstract}




\maketitle
\section{Introduction}

Online Q\&A forums such as Stack Overflow (SO) have become an integral part of modern programming workflow~\cite{interleaving_web, how_programmers_search, how_use_so_code, abdalkareem2017developers}. However, locating essential information in online posts can be time-consuming since there are often multiple relevant posts to consider, and some posts can be lengthy. This is confirmed by a recent survey with 72 professional software developers~\cite{answerbot}. Participants complained that sifting through many online posts was cognitively demanding and wished to get tool support for quickly navigating online posts.

Generating concise summaries of SO posts is a promising way to facilitate the navigation of SO posts~\cite{essential_sentences, answerbot, silva, huang2018api}. Nadi and Treude experimented with four approaches to capture the gist of a SO post by extracting essential sentences from it~\cite{essential_sentences}. They conducted a user study with 43 developers and confirmed that seeing essential sentences in a SO post indeed increased developers' confidence when assessing the relevance and quality of the post. However, they also found that selecting sentences only based on heuristics or information retrieval (IR) methods was not sufficient. Indeed, these approaches are inherently limited since they lack the flexibility of handling ambiguous or sophisticated narratives in natural language.

The Natural Language Processing (NLP) community has made significant progress in text summarization using deep learning (DL)~\cite{qa, bart, narayan2018ranking, dong2018banditsum, liu2019text}. 
Those DL-based approaches require to be trained with massive parallel corpora of text documents and summaries. For example,  the CNN/DailyMails dataset~\cite{dailymail} contains 286K pairs of news articles and human-written summaries.
However, no such large datasets exist for SO posts, except for a recent dataset named \dataset{}~\cite{kou2022sosum}. While \dataset{} contains manually curated summaries for \postnum{} SO answer posts, it is still much smaller compared with general-domain corpora such as CNN/DailyMails. Furthermore, no experiment has been done to prove that this dataset is sufficient to train a DL model for summarizing SO posts.

To bridge the gap, we propose a framework for SO post summarization, \name{} (\textsc{\underline{A}}utomated \textsc{\underline{S}}ummarization of \textsc{\underline{S}}tack \textsc{\underline{O}}ve\textsc{\underline{r}}flow pos\textsc{\underline{t}}). \name{} provides a novel supervised model {\sname} to account for the unique characteristics of SO posts. First, {\sname} uses a combination of BERT embeddings and domain-specific features to characterize sentences in a SO post. 
Second, since answers to different types of questions have different linguistic norms, we train a question classifier and separate models with the same BERT-based architecture for different types of questions. Third, to account for the uncertainty of the question classifier, the outputs of these models are then ensembled based on the probability distribution of the question classifier, as shown in Figure~\ref{fig:ensemble}.

Furthermore, since acquiring labeled training data is costly, \name{} also includes an indirect supervision method called {\uname}, which does not need to be trained with any labeled SO data.  {\uname} makes use of supervision signals from pre-trained
models in another domain, such as news articles, for SO post summarization. To handle domain shift issues, {\uname} uses a pre-trained Natural Language Inference (NLI) model to trace back to key sentences in the original SO post based on the initially generated summary. Compared with the initially generated summary, which may contain inaccurate terminologies and narratives due to domain shift, extracting aligned sentences from the original post can more accurately capture the gist of the post.

We evaluate both learning methods in \name{} with the comparison to a BERT-based text summarization model~\cite{liu2019fine}, which is fine-tuned with the same SO training data as {\sname}. We also select three heuristics-based methods and one unsupervised learning method from prior work~\cite{essential_sentences, robillard2015recommending, DBLP:journals/corr/abs-1109-2128} as baselines. 
We find that both {\sname} and {\uname} outperform all baselines by at least 13\% and 7\% in terms of the F1 score. This implies that only finetuning a general model with a relatively small dataset such as SOSum~\cite{kou2022sosum} is not sufficient. To improve training efficiency, it is necessary to account for the unique characteristics of SO posts. Furthermore, since {\uname} is not trained on any SO data, it achieves lower accuracy than {\sname}. However, when less than 20\% of the original training data are available, {\uname} achieves better summarization accuracy than {\sname}. This implies that indirect supervision remains a promising yet cheap alternative for building generalizable models for specific domains when there is a lack of labeled training data.

We conduct a qualitative user study with 12 participants to evaluate the quality of summaries generated by different techniques. Compared with the best baseline model~\cite{liu2019fine}, the majority of participants (76\%-85\%) preferred summaries generated by {\sname} or {\uname} in terms of usefulness, comprehensiveness, and conciseness. Participants found the summaries generated by {\sname} more concise and useful in practice, while they found the summaries generated by {\uname} more comprehensive and provided a better overview. Overall, neither {\sname} or {\uname} really triumph over each other according to user feedback.


In summary, this work makes the following contributions:
\begin{itemize}
    \item We design a new supervised model that generates concise and self-contained summaries of SO posts. This model accounts for the uniqueness of SO posts and achieves state-of-the-art accuracy in SO post summarization. 
    \item We propose a new indirect supervision method to further tackle the challenge of lacking labeled training data in SO post summarization. We demonstrate the feasibility of developing models with acceptable accuracy but with no cost of data labeling.  
    \item We conduct a comprehensive evaluation of the proposed learning methods with six comparison baselines, an ablation study, and a qualitative user study. 
\end{itemize}

The rest of the paper is organized as follows. Section II describes a motivating example for generating post summaries to facilitate the navigation of SO posts. Section III defines the SO post summarization task. Sections IV and V describe the supervised and indirect supervision models respectively. Section VI describes the evaluation design and setup. Section VII describes the evaluation results. Section VIII discusses the implications, threats to validity, and future work. Section IX describes the related work. Section X concludes this work. Section XI describes the data availability.

\section{motivating example}

\begin{figure*}[h]
    \centering
    \includegraphics[width=\linewidth]{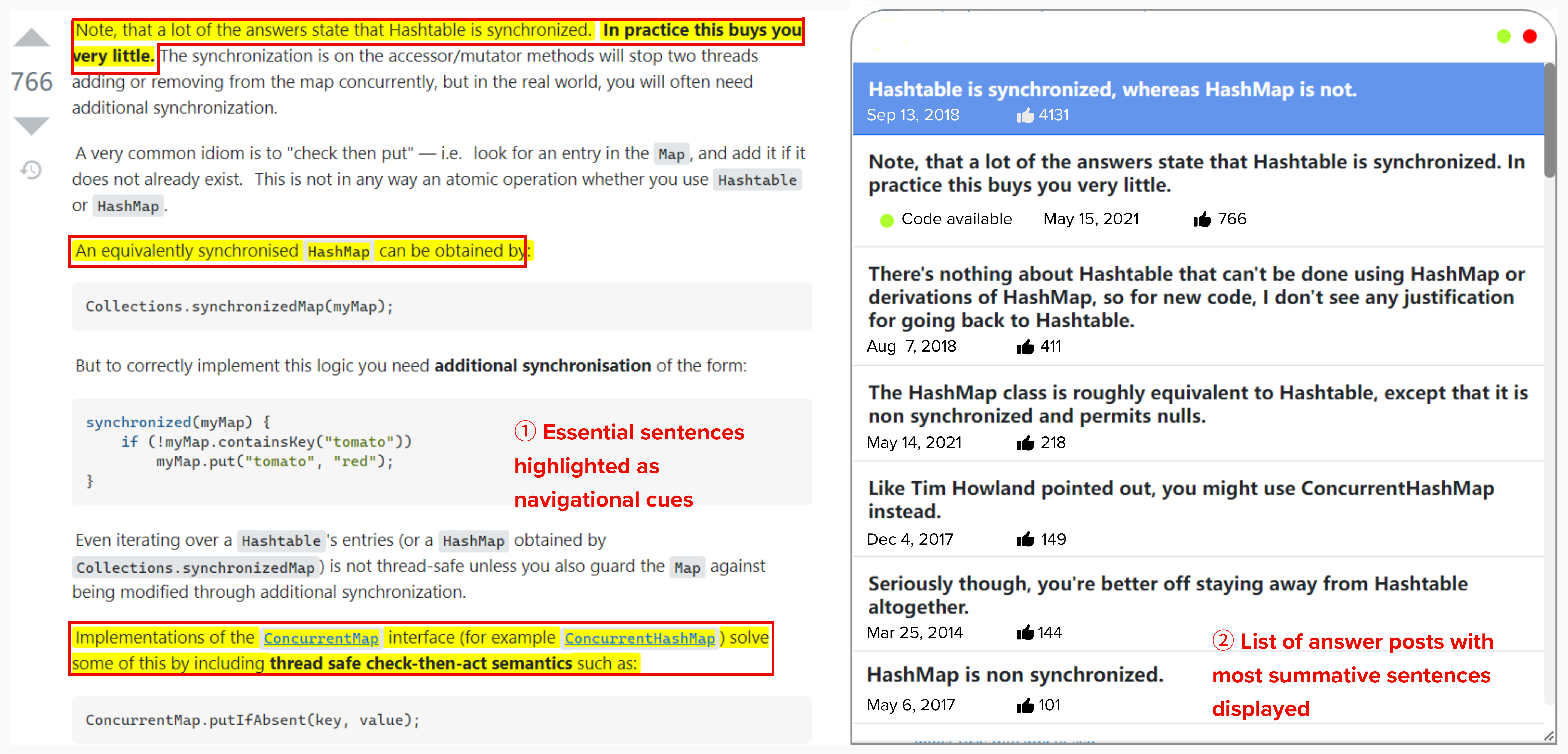}
    \caption{A Chrome extension built upon \name{}. On the left side, summative sentences are highlighted in yellow. On the right side, a navigation panel provides a  bird's-eye view of the thread by showing the first sentence in the summary of each answer.}
    \label{fig:soft_raw}
\end{figure*}

This section illustrates how summarizing SO posts helps developers quickly navigate SO posts and get an overview of various answers given by other developers. Suppose Alex is an Android developer and she wants to transfer key-value pairs between two layers of a Spring MVC framework. She knows both \texttt{HashMap} and \texttt{HashTable} can serve as the data structure for this task. Yet she is not sure about the pros and cons of these two APIs. So she searches online. 

The first search result from Google is a Stack Overflow question---\emph{``What are the differences between a HashMap and a HashTable in Java?''}\footnote{\url{https://stackoverflow.com/questions/40471}}
There are 35 answer posts to this question. Alex finds it time-consuming to read all of them. So she decides to first read the accepted answer. The accepted answer (Post 40878) points out three major differences between \texttt{HashMap} and \texttt{HashTable}: (1) \texttt{HashTable} is synchronized, whereas \texttt{HashMap} is not; (2) \texttt{HashTable} does not allow \texttt{null} keys or values; (3) \texttt{HashMap} has the flexibility to be replaced with \texttt{LinkedHashMap}. Alex finds this answer helpful, but she is not sure how comprehensive this answer is. 

Alex starts reading other highly voted answers to check if they include any important information not covered by the accepted answer. However, she finds herself submerged by the abundant information in those posts. Since many posts are lengthy with details that she does not care about, Alex spends most of her time locating helpful information in those posts. 
For example, Figure~\ref{fig:soft_raw} shows the second answer (Post 41454) in this thread. It is a long post with 4 code snippets. 
Yet the gist of it is a simple message---{\em the synchronization in \texttt{HashTable} is not sufficient,  and synchronized \texttt{HashMap} and \texttt{ConcurrentMap} are better choices}. Without any tool support, Alex has to read the entire post linearly to get this key message, which can take quite a few minutes. 

Like many other developers, Alex only reads a handful of answers and returns to her own code due to her limited time budget~\cite{brandt2009two, how_programmers_search, xia2017developers}. This inevitably makes her overlook answers that are not highly ranked but contain useful information that she is unaware of. For example, the 20th answer (Post 37031553) mentions that \texttt{HashMap} has $O(log(n))$ complexity and is faster than \texttt{HashTable}. None of the top five answers mentions this. 
If performance is a top concern to Alex, reading this post can help Alex make a more informed decision. However, without any tool support, Alex is unlikely to reach this deep in the thread practically.

\name{} helps Alex by automatically summarizing each answer post into concise text, so Alex can get a quick overview of many posts and prioritize which posts to spend more time on. In this way, she can make a more informed decision on which API to use. Specially, we build a Chrome extension for Stack Overflow, which highlights summative sentences from each post and renders a list of post summaries, as shown in Figure~\ref{fig:soft_raw}. By looking at the navigation panel, Alex quickly realizes that three answers mention \texttt{HashTable} is synchronized. By contrast, the other four answers suggest staying away from \texttt{HashTable}. 
Alex is interested in the second answer since its author expressed a strong opinion against \texttt{HashTable}. After clicking on it, Alex jumps to the corresponding answer, where three summative sentences have been highlighted as navigation cues. The first sentence points out that the synchronization in \texttt{HashTable} is not sufficient, while the second and third sentences propose synchronized \texttt{HashMap} and \texttt{ConcurrentMap} as alternatives. Alex feels that these highlighted sentences summarize the gist of the answer well, so she returns to the navigation panel to go deeper in the thread. As she scrolls down the panel, she notices the summary of the 20th answer---\emph{$O(log(n))$ for \texttt{HashMap} vs.~$O(n)$ in \texttt{HashTable}}. This indicates that \texttt{HashMap} has lower time complexity than \texttt{HashTable}. Alex dives into this answer for details, as performance is always a top concern for her. Without \name{}, this information would have been buried deep in the thread and unlikely to be discovered. Being aware of this information, Alex now feels more confident in making an informed decision between these two APIs.
\section{Task Definition}
We introduce the definition of \textit{Stack Overflow Post Summarization} as follows: Given a SO answer post consisting of $N$ sentences, the goal is to select a small set of sentences to form a succinct and self-contained summary. Essentially, this can be viewed as a contextualized sentence classification task where a sentence can either be in or not in the summary.

This task is also known as \textit{Extractive Summarization} (ES) in NLP. It is in contrast to another type of summarization called \textit{Abstractive Summarization} (AS). Instead of selection summative sentences from the original document, AS  generates new text to summarize the document~\cite{gupta2019abstractive}. Compared with ES, a unique challenge in AS is that distorted or fabricated text is likely to be introduced during text generation~\cite{huang2021factual, kryscinski2019evaluating}. Several studies have shown that factual inconsistency occurs in up to 30\% of abstractive summaries~\cite{cao2018faithful, kryscinski2019neural, goodrich2019assessing, falke2019ranking}. Furthermore, since most abstractive summarization models are pre-trained on 
news articles or general-domain corpora, such errors may become more prevalent when reusing those models on SO posts due to domain shift. 

In this work, we focus on extractive summarization as the first step towards SO post summarization while leaving the more challenging abstractive summarization task as future work. In Section~\ref{sec:supervised}, we first propose a supervised learning method with a novel model architecture tailored for SO posts. Then in Section~\ref{sec:indirect}, we propose an indirect supervision method that reuses pre-trained models from another domain while addressing domain shift issues via natural language inference. 

\section{Supervised Post Summarization}
\label{sec:supervised}

\begin{figure}[!t]
    \centering
    \includegraphics[width=\linewidth]{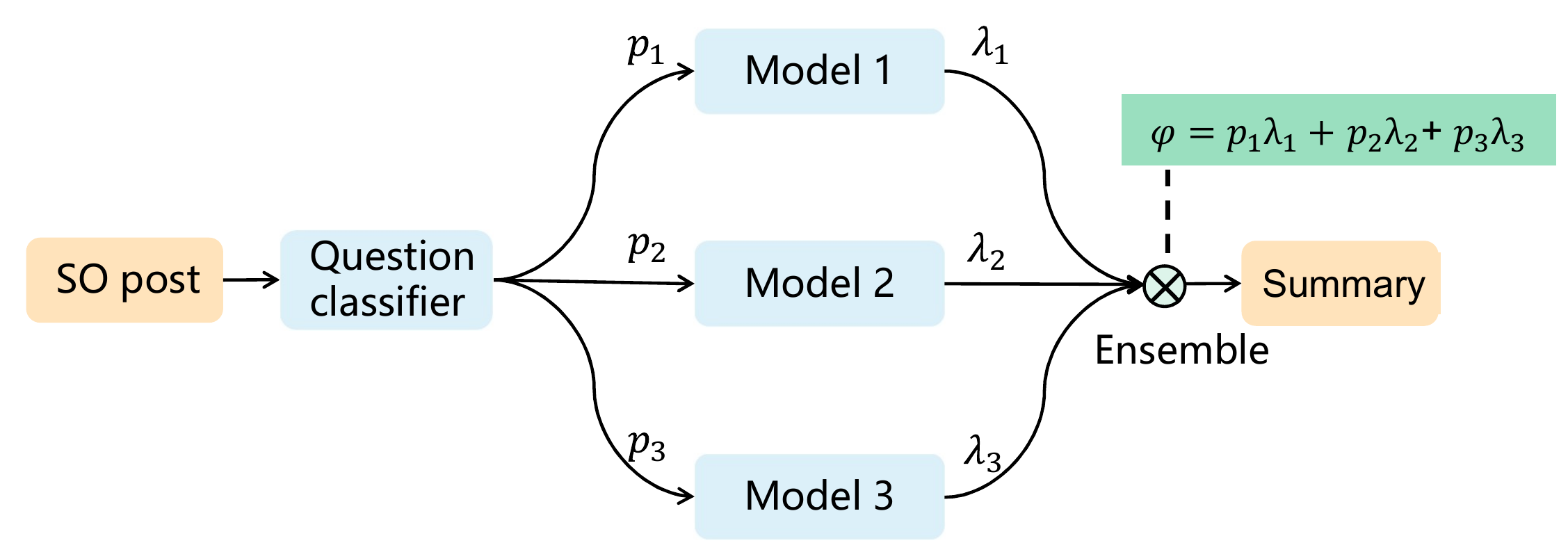}
    \caption{An overview of \sname{}}
    \label{fig:ensemble}
\end{figure}

\name{} includes a supervised learning method named \sname{}. As shown in Figure~\ref{fig:ensemble}, \sname{} takes three phases to summarize a SO answer post. Since answers to different types of questions follow different linguistic patterns, \sname{} first predicts the type of the SO question (Phase I). To account for the uncertainty of the question classifier, the answer post is fed into three sentence classification models separately, each of which is trained for one type of SO questions (Phase II). 
Finally, \sname{} ensembles the predictions of these models based on the likelihood of the question type to generate the final summary (Phase III).

\subsection{Question Classification}
While manually inspecting SO answer posts, we observed that answers to different kinds of questions have different linguistic forms. For example, answers to a how-to question often contain a step-by-step solution, while answers to a conceptual question often contain a definition or description of the concept. Based on this insight, we decide to first categorize SO posts based on their question types and then train separate post summarization models for different types of questions. We follow the SO question taxonomy from prior work~\cite{kou2022sosum, treude2011programmers, question_type, rosen2016mobile, allamanis2013and} and consider three representative types of SO questions---{\em how-to questions}, {\em conceptual questions}, and {\em bug fixing questions}. How-to questions ask for instructions for achieving a task, e.g., \emph{``how do I undo the most recent local commits in Git?''.} Conceptual questions ask for clarifications on a concept, e.g., \emph{``what are metaclasses in Python?''.} Bug fixing questions ask for solutions to fix some issues, e.g. \emph{``git is not working after macOS Update''.}

We develop a Support Vector Machine (SVM) classifier to categorize SO posts based on their question titles. 
We train it with a combination of 506 classified SO questions from \dataset{}~\cite{kou2022sosum} and 365 new questions.  In total, our dataset includes \howtonum{} how-to questions, \whatisnum{} conceptual questions, and \debugnum{} bug-fixing questions.  {We choose these three types of questions, since they are the most common question types, covering 77\% of SO questions based on prior work~\cite{question_type}. We discard questions that did not belong to the three types of questions when curating the dataset.} 

 {To select and label the additional 365 questions, the first author first ranked all SO questions by view count in descending order. Then, he inspected these questions and manually classified them based on the types until he found 365 questions belonging to one of the three types of questions under investigation. Then, the first author and another undergraduate student independently labeled the summative sentences in the answer posts of these questions following the labeling heuristics described in the SOSum paper~\cite{kou2022sosum}. In total, they labeled 785 answer posts under these 365 questions. The Cohen's Kappa score before the discussion is 0.72, which implies a substantial agreement~\cite{mchugh2012interrater}. The two labelers met to discuss their labels and resolved all disagreements. This labeling process took about 73 man-hours.}
 
Despite the simplicity of SVM, this classifier achieves reasonable accuracy (78\%) with 8:1:1 train/dev/test data split and 10-fold validation.
To further account for the potential misclassification of the SVM classier, {\sname} adopts an ensemble mechanism that incorporates the probability distribution of different types of questions predicted by the SVM classifier (Section~\ref{sec:ensemble}). An ablation study confirms the benefit of incorporating question classification into {\sname}, improving its F1-score  by 9\% (Section~\ref{sec:ablation}).

\subsection{Summative Sentence Identification}
We train separate models to identify summative sentences for the three representative types of SO questions. These models share the same model architecture but are trained on answer posts to different types of questions to capture unique linguistic norms and writing styles of each question category. We explain the model architecture below.

\begin{figure}[!t]
    \centering
    \includegraphics[width=1\linewidth]{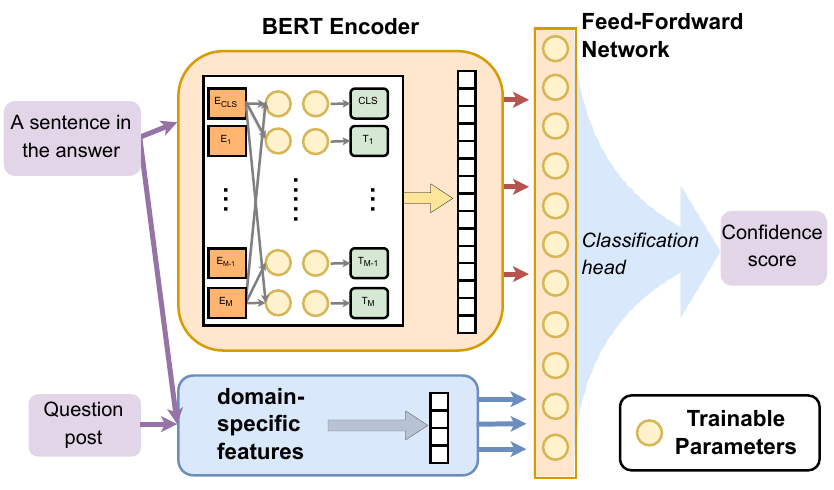}
    \caption{The architecture of the sentence classifier}
    \label{fig:supervised}
\end{figure}

Figure~\ref{fig:supervised} gives an overview of the model architecture. \sname{} performs hybrid learning by combining the semantic representation from a BERT model and a multifaceted set of domain-specific features. Each sentence in the given post is encoded into a vector embedding, which is then fed into a feedforward neural network (FNN) to decide whether the sentence is a summative sentence.  

In this work, we use a BERT model that is fine-tuned with 152 million sentences from Stack Overflow~\cite{tabassum2020code}. Given a sentence from a SO post, \sname{} computes its embedding from BERT by averaging the embedding vectors of all tokens in the sentence.
The incorporation of deep contextualized sentence embeddings helps {\sname} capture the semantics of a given sentence. This is confirmed by the ablation study in Section~\ref{sec:ablation}, where the inclusion of BERT increases the F1 score of our model by 17\%. 

While the contextualized embeddings from BERT are effective in capturing sentence semantics, they are not sufficient to capture specific features, such as bold text, which human readers often rely on to identify summative sentences. Therefore, we further design a set of domain-specific features to capture the semantic, structural, and stylistic information cues in SO posts. We elaborate on these features below.

First, we design five \textbf{semantic features} to select summative sentences based on their content.
\begin{itemize}
    \item \emph{Entity Overlap.} If a sentence mentions a  software entity that is also mentioned in the question title or in a SO tag of the question, this sentence may provide some directly relevant information for the question. In this work, we use SEthesaurus~\cite{chen2017synonym} to identify software entities in a sentence. Let $E_q$ be the combination of software entities in the question title and the SO tags in the question. Let $E_{s}$ be the set of software entities mentioned in a sentence in the answer post. The entity overlap is computed as $|E_q|\cap|E_s|/|E_q|$.
    \item \emph{Comparative Adjective.} This feature captures whether a sentence contains a comparative adjective. Sentences containing a comparative adjective often contain information that helps readers compare two APIs or two solutions, e.g.,  \emph{``the stack is faster because all free memory is always contiguous.''}
    \item \emph{Superlative Adjective.} This feature captures whether a sentence contains a superlative adjective. Similar to comparative adjectives, sentences with superlative adjectives often indicate answerers' strong inclination for or against an approach, API, or bug fix, e.g., \emph{``there is no doubt that application/json is the best MIME type for a JSON response.''}
    
    \item \emph{Imperative Sentence.} This feature captures whether a sentence is an imperative sentence. Imperative sentences often contain instructions to accomplish a programming task or to fix a bug, e.g., \emph{``use git revert commit-id.''} 
    
    \item \emph{Linguistic Patterns.} As we label SO posts for training \sname{}, we summarize 19 phrases that may imply important information in an answer (Table~\ref{table:semantic_features}).  {Specifically, the first author randomly sampled 100 posts from the labeled dataset and manually analyzed the summative sentences of these posts. He identified an initial set of common phrases that were shared across multiple sentences. He then applied these patterns back to those sentences to measure the coverage and iteratively refined the patterns. Similar procedures have been adopted in prior work to identify linguistic patterns~\cite{robillard2015recommending, api, answerbot}.} Each pattern corresponds to a dimension in the sentence embedding. If a linguistic pattern is matched, then the corresponding dimension is set to 1, otherwise 0. 
\end{itemize}

Second, we design two \textbf{structural features} to select summative sentences based on their relations with other sentences and codes in the post.
\begin{itemize}
    \item \emph{Sentence Position.} This feature captures the position of a sentence in a given post. This feature is designed based on our observation that the leading sentences in many SO posts can serve as a good summary of the post. 
    \item \emph{Code Adjacency.} This feature indicates whether a sentence is around a code snippet. It is designed based on our observation that sentences around a code snippet enclosed by a \texttt{pre} tag often contain information cues for a programming task solution or a bug fix.
\end{itemize}

Third, we design three \textbf{stylistic features} to capture formatting styles that are used to highlight important information in a SO post.
\begin{itemize}
  \item \emph{Inline Code.} This feature indicates whether a sentence contains an inlined code fragment. This feature is designed based on our observation that, for how-to questions and bug-fixing questions, answerers sometimes suggest an alternative API, pinpoint a problematic piece of code, or provide a short code fragment as a solution. 
  \item \emph{Bold Text.} This feature indicates whether a sentence contains bold text. In SO posts, answerers often highlight important terms or statements in bold to draw attention from readers. 
  \item \emph{Step in a List.} This feature indicates whether a sentence is the first sentence of an item in a bulleted or numbered list. This feature is designed based on our observation that for how-to questions, many answerers typically provide a list of steps to accomplish a task. 
\end{itemize}

The ablation study (Section~\ref{sec:ablation}) confirms the usefulness of these domain-specific features. Specifically, including these features leads to an increase of 6\% in the F1 score.  {We also conducted an experiment to measure the contribution of each feature by removing each of them and evaluating the accuracy degradation. The results are included in the Supplementary Material.}

\begin{table}[]
\caption{Linguistic patterns for sentences that may contain important information}
\begin{center}
\resizebox{0.45\textwidth}{!}{
\renewcommand{\arraystretch}{1}
\SetTblrInner{rowsep=1pt}
  \begin{tblr}{
      width = 0.40\textwidth,
      colspec={Q[c,Gray]|l|Q[c,Gray]|l},
      row{1, 2, 3, 4, 5, 6, 7, 8, 9, 10, 11} = {font = \scriptsize},
      rowspec={Q[Gray]},
      stretch=0
    }
\toprule
\textbf{No.} & \\textbf{Phrase} & \textbf{No.} & \textbf{Phrase} \\ \hline
1   &  {However, ...}       & 11  &   {In practice, ...}       \\
2   &    {First, ...}     & 12  &     {In fact, ...}     \\
3   &    {In short, ...}       & 13  &     {Otherwise, ...}     \\
4   &    {In this case, ...}       & 14  &   {If you care, ...}       \\
5   &    {In general, ...}      & 15  &    {In contrast, ...}      \\
6   &      {Finally, ...}    & 16  &     {On the other hand, ...}      \\
7   &    {Then, ...}       & 17  &     {Below is ...}     \\
8   &    {Alternatively, ...}      & 18  &    {Additionally, ...}     \\
9   &    {In other words, ...}       & 19  &     {Furthermore, ...}     \\
10  &    {In addition, ...}       &   &    \\
\bottomrule
\end{tblr}}
\label{table:semantic_features}
\end{center}
\end{table}

\subsection{Ensemble Inference}
\label{sec:ensemble}

To account for the uncertainty of question classification in Phase I, we design an ensemble mechanism that merges the predictions from the three sentence classifiers to make the final decision about whether a sentence should be included in the summary. 

The ensemble mechanism takes two input: (1) the softmax probabilities of $k$ categories ($k=3$ in this case) predicted by the question classifier, $p_1, ..., p_k$; (2) the probabilities of a sentence being a summative sentence from each sentence classifier, $\lambda_1, ..., \lambda_k$. The final score $\phi$ of a sentence is defined as:
$\phi = \sum^k_{i=1}p_i\lambda_i$. 
If a sentence has a final score greater than a threshold $\theta$, it is selected as a summative sentence. We experimentally determine $\theta$ to be $0.5$ on a validation set of 303 posts (10\% of the dataset) with the objective of maximizing the F-1 score. After every sentence in an answer post is classified, \name{} collects all the selected sentences and outputs them as an extractive summary for the answer post. 

\section{Post Summarization via Indirect Supervision}
\label{sec:indirect}

\begin{figure}[h]
    \centering
    \includegraphics[width=0.9\linewidth]{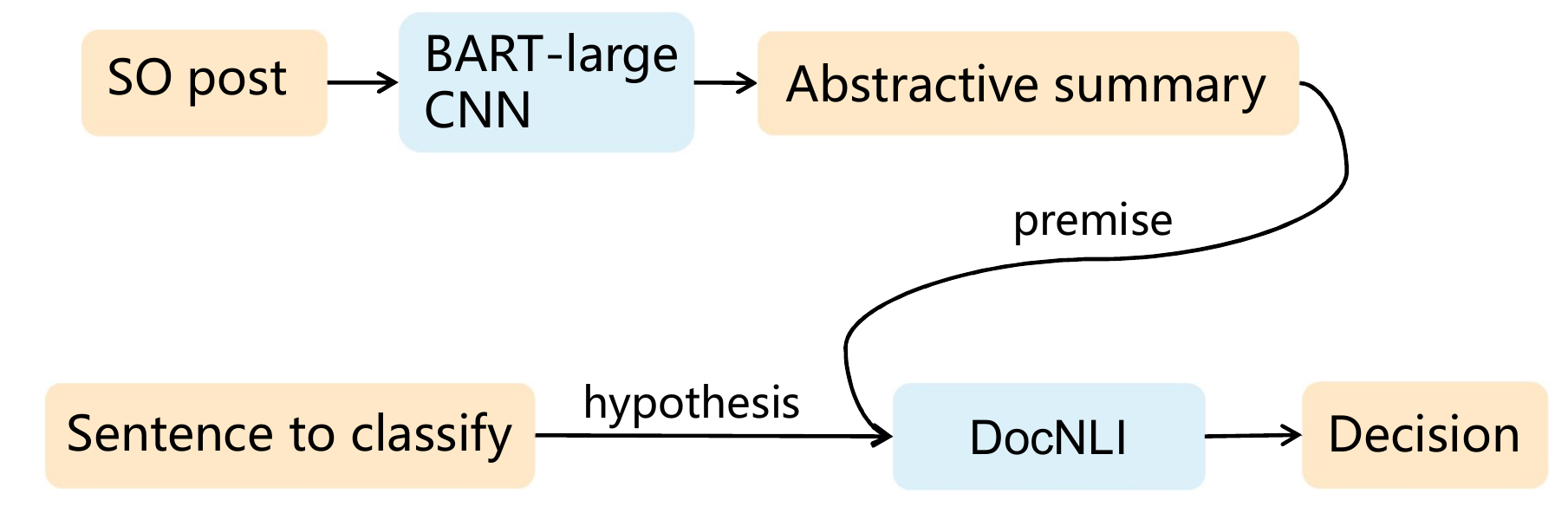}
    \caption{An overview of \uname{}}
    \label{fig:indirect}
\end{figure}

While supervised learning can achieve superior performance,
obtaining a large amount of labeled data is often costly, especially in specific domains such as software engineering.
Therefore, we propose an indirect supervision approach, \uname{}, to overcome this limit.
Instead of acquiring labeled data 
for direct supervision, \uname{} uses supervision signals from pre-trained models in another domain, such as news article summarization. 
To address the challenge of data shift in cross-domain transfer, we use a pre-trained Natural Language Inference (NLI) model to select summative sentences in the original post based on the summary generated by the pre-trained text summarization model. Figure~\ref{fig:indirect} provides an overview of our approach. 

In this work, we use BART-large-CNN~\cite{barthuggingface}
as the pre-trained text summarization model. Unlike our supervised model in Section~\ref{sec:supervised}, BART-large-CNN is an abstractive summarization model. In other words, BART-large-CNN generates a summary in its own words via an autoregressive decoding process, rather than selecting summative sentences in the input document. Specifically, BART-large-CNN is based on BART~\cite{bart}, a denoised auto-encoder that encodes an input document to a high-dimensional embedding for text reconstruction. In BART-large-CNN, the initial BART model is fine-tuned to generate text summaries using the CNN/DailyMail corpus~\cite{dailymail}, which contains 300K news articles and their human-written summaries. 
In this work, we select BART-large-CNN,  since it is 
a representative and strong summarization model that
outperforms other transformer-based models such as~\cite{liu2019text} by up to 6 points on CNN/DailyMail in ROUGE-based metrics~\cite{bart}. 


\begin{table}
  \caption{An abstractive summary generated by BART-large-CNN and the summary generated by our approach}
  \label{table:bart_example}
  \centering
  \normalsize
\renewcommand{\arraystretch}{1.2}
\resizebox{0.5\textwidth}{!}{
\begin{tabular}{p{0.50\textwidth}}
\hline
\rowcolor{Gray}
\textsf{Original post [\href{https://stackoverflow.com/questions/438095/custom-windows-workflow-activity-that-executes-an-asynchronous-operation-redon/439105\#439105}{Post 438095}]}                                                  \\\hline                                                     
There is a potential problem here with regard to workflow persistence.

\ \ \ \ If you create long running workflows that are persisted in a database, these workflows are not reloaded into memory until there is some external event that reloads them. As there they are responsible for triggering the event themselves but cannot until they are reloaded. And we have a catch 22 :-(.

\ \ \ \ \hlgreen{The proper way to do this is using an external service.} And while this might feel like dividing the code into two places it really isn't. The reason is that the workflow is responsible for the big picture, IE what should be done. And the runtime service is responsible for the actual implementation or how it should be done. \hlgreen{That way you can change the how without changing the why and when part.} \textbf{[Source of the ineloquent sentence in BART-large-CNN's summary]}
\\\hline
\rowcolor{Gray}
\textsf{Abstractive summary generated by BART-large-CNN} \\ \hline
There is a potential problem with regard to workflow persistence. If you create long running workflows that are persisted in a database they will not be able to restart until there is an external event reloading them. \hlred{The proper way of doing this is using an external service which can change the how part without changing what it should be done or when it's better than having two separate processes for different tasks as each one does their own thing but doesn't agree on exactly who must carry out its function at any given time.} \textbf{[broken English]}
\\ 
\hline
\rowcolor{Gray}
\textsf{Summary generated by our indirect supervision approach} \\ \hline
There is a potential problem here with regard to workflow persistence. The proper way to do this is using an external service. That way you can change the how without changing the why and when part.
\\ 
\hline
\end{tabular}
}
\end{table}

While BART-large-CNN is demonstrated effective on news article summarization, it suffers from 
domain-shifting issues when applied to a different domain, such as SO post summarization. Furthermore, recent studies have shown that abstractive summarization models may introduce factual inconsistencies in generated text compared with original text~\cite{huang2021factual, kryscinski2019evaluating}. Table~\ref{table:bart_example} shows an abstractive summary generated by BART-large-CNN. The last sentence in it (highlighted in red) is exceptionally long and contains broken English with some hallucinated information.

Despite the broken English, this sentence is not fully made up by BART-large-CNN. It is based on two sentences in the original post (highlighted in green). In fact, these two sentences provide a good summary of the last paragraph in the original post. Though BART-large-CNN recognizes them, it fails to generate an eloquent summary due to domain shift.

To address this issue, we propose associating abstractive summaries to the original post's corresponding sentences using a pre-trained Natural Language Inference (NLI) model. The task of NLI~\cite{dagan2005pascal} studies to what extent a 
premise entails, contradicts, or remains neutral with a 
hypothesis. For example, suppose we have a hypothesis sentence, ``{\em a kid ate a fruit}'', and a premise sentence, ``{\em a boy ate an apple}''. An NLI model will predict that the premise entails the hypothesis. However, if the hypothesis is changed to ``{\em a kid ate a banana}'', the prediction will becomes to neutral or contradiction.

We use a pre-trained NLI model to decide which sentence in the original SO post is entailed by the abstractive summary and thus should be included in the final summary. Specifically, we use a RoBERTa model that is pre-trained on DocNLI~\cite{yin2021docnli}, a document-level NLI dataset. DocNLI covers multiple text genres, such as news, fiction, and conversations, with multi-sentence (i.e., document) premises and single-sentence hypotheses. Given a summary generated by BART-large-CNN, our approach checks the logical relationship between the summary (i.e., the premise) and every sentence in the original post (i.e., the hypotheses). The DocNLI model will then produce a probability distribution over the three relationships---{\em entail}, {\em contradict}, and  {\em neutral}. Our approach selects a sentence as part of the final summary if the entailment probability is the higher than the other two probabilities. 

\section{Evaluation}
\label{sec:expr}
We conduct both quantitative experiments and user studies to answer the following four research questions:
\begin{itemize}
    \item RQ1: How effective is our supervised model, \sname{}, in SO post summarization?
    \item RQ2: To what extent does each component in \sname{} contribute to its effectiveness?
    \item RQ3: How does our indirectly supervised model, \uname{}, compare to our surprised model?
    \item RQ4: How do real programmers perceive the summaries given by our models? 
\end{itemize}

\subsection{Experiment Setup}
In addition to the 2,278 labeled posts from \dataset{}~\cite{kou2022sosum}, we further manually labeled 785 answer posts following the same labeling procedure as described in~\cite{kou2022sosum}. The labeling process is described in Section~\ref{sec:supervised}. We use these 3,063 posts for training and evaluation. These posts are from 785 SO questions, including 254 {\em how-to} questions, 322 {\em conceptual} questions, and 209 {\em bug-fixing} questions. 

 {We empirically decided the model structure and hyperparameters for the classification head of \sname{}. Specifically, we experimented with different kinds of models, as shown in  Table~\ref{table:classification_heads}. We chose feedforward neural network (FNN) since it performed the best. We also experimented with different numbers of hidden layers in the FNN and eventually chose 1 hidden layer, as shown in Table~\ref{table:number_of_layers}.}

\begin{table}[h]
    \caption{Model Accuracy with Different Types of Classification Heads}
        \label{table:classification_heads}
\begin{center}
\resizebox{0.65\linewidth}{!}{
\begin{tabular}{ccc}

\hline
\multicolumn{1}{|l|}{} & \multicolumn{1}{c|}{\textbf{F1}} & \multicolumn{1}{c|}{\textbf{$\Delta$}}\\
\hline\hline
\multicolumn{1}{|l|}{\textbf{{Feedforward NN}}} & \multicolumn{1}{c|}{\textbf{0.71}} & \multicolumn{1}{c|}{\textbf{-}}\\ \hline
\multicolumn{1}{|l|}{Random forest} & \multicolumn{1}{c|}{{0.56}} & \multicolumn{1}{c|}{{-0.15}} \\
\multicolumn{1}{|l|}{Decision tree} & \multicolumn{1}{c|}{0.54} & \multicolumn{1}{c|}{-0.17} \\
\multicolumn{1}{|l|}{Linear regression} & \multicolumn{1}{c|}{0.65} & \multicolumn{1}{c|}{-0.06} \\
\multicolumn{1}{|l|}{Logistic regression} & \multicolumn{1}{c|}{{0.63}} & \multicolumn{1}{c|}{{-0.08}} \\
\multicolumn{1}{|l|}{Ada boost} & \multicolumn{1}{c|}{{0.59}} & \multicolumn{1}{c|}{{-0.12}} \\
\multicolumn{1}{|l|}{Naive Bayes Classifier} & \multicolumn{1}{c|}{0.62} & \multicolumn{1}{c|}{-0.09}\\
\hline
\end{tabular}}
\end{center}
\end{table}

\begin{table}[hbt!]
    \caption{Model Accuracy with Different Numbers of Hidden Layers in the FNN}
        \label{table:number_of_layers}
\begin{center}
\resizebox{0.75\linewidth}{!}{
\begin{tabular}{cccc}

\hline
\multicolumn{1}{|l|}{\# hidden layers} & \multicolumn{1}{c|}{\textbf{Precision}} & \multicolumn{1}{c|}{\textbf{Recall}} & \multicolumn{1}{c|}{\textbf{F1}} \\
\hline\hline
\multicolumn{1}{|l|}{1} & \multicolumn{1}{c|}{\textbf{0.73}} & \multicolumn{1}{c|}{\textbf{0.69}} & \multicolumn{1}{c|}{\textbf{0.71}} \\ \hline
\multicolumn{1}{|l|}{2} & \multicolumn{1}{c|}{0.69} & \multicolumn{1}{c|}{0.71} & \multicolumn{1}{c|}{0.70} \\
\multicolumn{1}{|l|}{3} & \multicolumn{1}{c|}{0.73} & \multicolumn{1}{c|}{0.71} & \multicolumn{1}{c|}{0.72} \\
\multicolumn{1}{|l|}{4} & \multicolumn{1}{c|}{0.73} & \multicolumn{1}{c|}{0.70} & \multicolumn{1}{c|}{0.71} \\
\hline
\end{tabular}}
\end{center}
\end{table}

To train \sname{}, for each type of questions, we randomly split the corresponding answer posts into training, development, and test sets with a split ratio of 8:1:1. Then, we train a summative sentence identification model for answers to each type of question following the model architecture in Figure~\ref{fig:supervised}.  {During training and testing, an answer post to be summarized is broken down into individual sentences. In total, we have 14,165 sentences for the 2,424 answer posts in the training set. Each model was trained with a batch size of 512 sentences in 150 epochs. In each batch, each sentence is fed to \sname{} one by one with the corresponding question title.} We use the Adam optimizer~\cite{kingma2014adam} for training and the learning rate is set to $1e^{-5}$.  {Training \sname{} took 3 hours on a single GPU (NVIDIA GeForce GT 1030).}

To design an ablation study to answer RQ2, we create four variants of \sname{} with one key component removed in each variant. The four key components are: (1) the BERT embeddings, (2) the domain-specific features, (3) the question classifier, (4) the ensemble mechanism. Specifically, when the question classifier is ablated, \sname{} uses a sentence classifier trained on all types of questions to predict the final score of a sentence. 
When the ensemble mechanism is ablated, \sname{} first predicts the question type of a given post and then only uses the sentence classifier for the predicted question type to identify summative sentences, rather than using all three classifiers. 

To answer RQ3, we first compare \uname{} against \sname{} on {\em a full training setting} where we train \sname{} with the entire training set. Then, we compare them in different {\em low-resource settings} where only a subset of the original training data is available. In those low-resource settings, we randomly select SO posts from the original dataset to make subsets of training data. Since \uname{} does not require any training data, the low-resource baselines are created to make a fair comparison between \sname{} and \uname{} in conditions where training data are scarce.

\subsection{Compared Baselines}
We select a state-of-the-art 
extractive summarization model on the general domain~\cite{liu2019fine} and also fine-tune it on the same training set of 2,424 SO posts as \sname{}. Furthermore, we select three heuristics-based methods and one unsupervised method that can perform sentence-level summarization from prior work~\cite{essential_sentences, robillard2015recommending, DBLP:journals/corr/abs-1109-2128}. These four methods have also been experimented in \cite{essential_sentences}. 

\begin{enumerate}[label={(\arabic*)}]
  \item \liu{}~\cite{liu2019fine} is an extractive summarization model that first uses BERT to encode sentences and then uses a transformer to select summative sentences~\cite{liu2019fine}. It outperforms several previous techniques~\cite{narayan2018ranking, zhou2018neural, see2017get, celikyilmaz2018deep} on two popular text summarization datasets---NYT~\cite{sandhaus2008new} and CNN/DailyMail~\cite{dailymail}. In this experiment, we use the checkpoint of \liu{} that has the best performance on CNN/DailyMail.
  
  \item \textit{\liu{} (fine-tuned)}~\cite{liu2019fine} is a fine-tuned version of \liu{}. It is fine-tuned with the training data of \sname{}, including 2,424 SO posts and their summaries.
  
  \item \emph{wordpattern} identifies essential sentences in a SO post using a set of 360 word patterns. These patterns are initially designed by Robillard and Chhetri~\cite{robillard2015recommending} to identify sentences containing indispensable knowledge in API documentations.
  
  \item \emph{simpleif} is a technique proposed by Nadi and Treude~\cite{essential_sentences}. It is designed based on the insight that essential sentences may contain contextual information expressed in the form of conditions. Thus, \emph{simpleif} identifies all sentences that have the word ``if'' in them as essential sentences. 
  
  \item \emph{contextif} is another technique proposed by Nadi and Treude~\cite{essential_sentences}. It uses a set of heuristics to identify essential sentences that carry technical context and are useful.

  \item \emph{lexrank} is a commonly used unsupervised text summarization approach~\cite{DBLP:journals/corr/abs-1109-2128}. It uses a stochastic graph-based method to compute the relative importance of sentences in a document and generates an extractive summary by selecting the top $k$ sentences. We use the default $k$ value, 5, in an open-source implementation of \emph{lexrank}~\cite{lexrank}.
  
\end{enumerate}

We do not compare with paragraph-level summarization techniques such as AnswerBot~\cite{answerbot} and CraSolver~\cite{wang2021automatic}, since it is not a head-to-head comparison. Take AnswerBot as an example.  {First, the problem setting is different. AnswerBot summarizes multiple SO threads, while ASSORT summarizes key points in a single post. Second, even if we adapt AnswerBot to only summarize a single post, AnswerBot can only produce a summary by selecting paragraphs. By contrast, ASSORT produces a more fine-grained summary by selecting sentences. Thus, AnswerBot always generates longer and more coarse-grained summarizations than \name{}, leading to low precision on the benchmark.}

\subsection{Evaluation Metrics}
We use three metrics---precision, recall, and F1 to measure the effectiveness of our methods and the comparison baselines in SO post summarization. Each metric is calculated at the sentence level. Given a set of summative sentences in a set of SO posts $G$, let $M$ be the set of summative sentences selected by an extractive summarization model. The precision of the model is calculated as ${|G\cap M|}/{|M|}$. And the recall of the model is defined as ${|G\cap M|}/{|G|}$. Furthermore, we measure the F1 score, which combines the precision and recall of a model into a single metric by taking their harmonic mean.

We make sure both \sname{} and \uname{} and all the baselines are evaluated with the same test set (i.e., 304 posts and their summaries) to make the comparison fair. A 10-fold validation is performed when calculating the metrics for our approaches and all the baselines.

\subsection{User Study Design}
\label{sec:design}

To answer RQ4, we conduct a within-subjects user study to evaluate the summary quality. We recruit 12 graduate students through the department mailing list from an R1 university. Participants have an average of four years of programming experience.
We randomly selected 40 answer posts from our dataset, including 13 \emph{how-to questions}, 13 \emph{conceptual questions}, and 14 \emph{bug-fixing questions}. 
 {These tasks were randomly sampled from the test set while ensuring a balanced number of answers in each category.}
In each user study, we select 10 out of the 40 answer posts and ask the participant to review their summaries. We counterbalance the post assignment so that each post is evaluated by three different participants.  

For each answer post, participants first report their expertise of the concepts in the question on a 7-point scale.  1 means ``Haven't even heard of it before'' and 7 means ``I am an expert''. Then, they will be provided with the question post, the answer post, and the summaries generated by \sname{}, \uname{}, and {\em BertSum (fine-tuned)}. Specifically, we select {\em BertSum (fine-tuned)} as our baseline in the user study since it performs the best among all six baselines in the quantitative experiment.  {The participants first read the question to understand the context and then read the answer post to understand the content to be summarized. The participants are then asked to read all three summaries and evaluate the quality of these summaries by answering the following five multiple-choice questions.}

\begin{enumerate}[label={(\arabic*)}]
    \item \textit{Which summary provides the most helpful information?}
    \item \textit{Which summary provides the best overview of the post?}
    \item \textit{Which summary provides the most comprehensive information?}
    \item \textit{Which summary is the most concise without being incomplete?}
    \item \textit{Which summary do you prefer to see in practice?}
\end{enumerate}

 {To mitigate bias, the order of summaries is randomized to mitigate bias and we also do not reveal which model generated which summary.} 
At the end of the user study, participants answer several open-ended questions about whether they wish to see SO post summaries when browsing SO and what kinds of characteristics an ideal SO summary should possess.

\section{Results}
\subsection{RQ1: SO Post Summarization Accuracy}
\label{sec:accuracy}
Table~\ref{table:rq1} shows the summarization accuracy of \sname{} and the baselines after 10-fold validation is performed. \sname{} achieves the best results in all three metrics. Furthermore, the domain shift from general text corpora to the SO post corpus is non-trivial. The original \liu{} model, which is trained on news articles and their summaries from CNNDailymails, only achieves an F1 score of 53\%. While fine-tuning \textit{\liu{}} with 2,424 labeled SO posts increases the F1 score from 53\% to 58\%, it is still 13\% below \sname{}. Given that \liu{} also uses BERT for sentence encoding, this result implies that directly reusing a pre-trained model, even with finetuning, is not an optimal solution. Incorporating domain-specific features and ensembling based on question types are necessary to improve the effectiveness of SO post summarization. 

\setlength\doublerulesep{0.1cm} 
\begin{table}[]
    \caption{Comparison of {\sname} and baselines}
        \label{table:rq1}
\begin{center}
\resizebox{0.80\linewidth}{!}{
\begin{tabular}{cccc}

\hline
\multicolumn{1}{|l|}{} & \multicolumn{1}{c|}{\textbf{Precision}} & \multicolumn{1}{c|}{\textbf{Recall}} & \multicolumn{1}{c|}{\textbf{F1}} \\
\hline\hline
\multicolumn{1}{|l|}{\textbf{{\sname}} $\filledstar$} & \multicolumn{1}{c|}{\textbf{0.73}} & \multicolumn{1}{c|}{\textbf{0.69}} & \multicolumn{1}{c|}{\textbf{0.71}} \\
\multicolumn{1}{|l|}{\liu{} $\filledstar$} & \multicolumn{1}{c|}{0.47} & \multicolumn{1}{c|}{0.60} & \multicolumn{1}{c|}{0.53} \\
\multicolumn{1}{|l|}{{\em BertSum (fine-tuned)} $\filledstar$} & \multicolumn{1}{c|}{0.51} & \multicolumn{1}{c|}{0.67} & \multicolumn{1}{c|}{0.58} \\
\multicolumn{1}{|l|}{wordpattern $\smalldiamond$} & \multicolumn{1}{c|}{0.40} & \multicolumn{1}{c|}{0.03} & \multicolumn{1}{c|}{0.06} \\
\multicolumn{1}{|l|}{simpleif	$\smalldiamond$} & \multicolumn{1}{c|}{0.39} & \multicolumn{1}{c|}{0.15} & \multicolumn{1}{c|}{0.21} \\
\multicolumn{1}{|l|}{contextif	$\smalldiamond$} & \multicolumn{1}{c|}{0.39} & \multicolumn{1}{c|}{0.15} & \multicolumn{1}{c|}{0.22} \\
\multicolumn{1}{|l|}{lexrank	$\smallstar$} & \multicolumn{1}{c|}{0.61} & \multicolumn{1}{c|}{0.45} & \multicolumn{1}{c|}{0.52} \\
\hline
\multicolumn{4}{c}{\footnotesize $\filledstar$: DL-based, $\smalldiamond$: heuristics-based, $\smallstar$: unsupervised}
\end{tabular}}

\end{center}
\end{table}

\subsection{RQ2: Ablation Study} 
\label{sec:ablation}
Table~\ref{table:ablation_study} shows the ablation study results. On the one hand, ablating the BERT embedding leads to the largest accuracy degradation, 17\% in the F1 score. This indicates that incorporating deep contextualized embeddings from a language model is critical for a domain-specific task such as SO post summarization. On the other hand, only including BERT is also not sufficient. Removing each of the other three components, which are specifically designed to account for the unique characteristics of SO posts, leads to a non-trivial decrease in the F1 score.
Specifically, the removal of domain-specific features decreases the F1 score by 6\%. This indicates that deep contextualized embeddings alone cannot cover the unique structural and linguistic patterns that distinguish SO posts from general text data. 
Removing the question classifier decreases the F1 score by 9\%. This indicates that accounting for different linguistic norms in answers to different types of questions indeed helps. Removing the ensemble mechanism leads to a decrease of 4\% in the F1 score. This indicates that the ensemble mechanism can help to alleviate the influence of question classification errors.

\setlength\doublerulesep{0.1cm} 
\begin{table}[]
    \caption{Contribution of each component in \sname{}}
        \label{table:ablation_study}
\begin{center}
\resizebox{0.85\linewidth}{!}{
\begin{tabular}{cccc}

\hline
\multicolumn{1}{|l|}{} & \multicolumn{1}{c|}{\textbf{Precision}} & \multicolumn{1}{c|}{\textbf{Recall}} & \multicolumn{1}{c|}{\textbf{F1}} \\
\hline\hline
\multicolumn{1}{|l|}{\textbf{{\sname}}} & \multicolumn{1}{c|}{\textbf{0.73}} & \multicolumn{1}{c|}{\textbf{0.69}} & \multicolumn{1}{c|}{\textbf{0.71}} \\
\multicolumn{1}{|l|}{\textemdash w/o BERT} & \multicolumn{1}{c|}{0.61} & \multicolumn{1}{c|}{0.48} & \multicolumn{1}{c|}{0.54} \\
\multicolumn{1}{|l|}{\textemdash w/o Domain-specific features} & \multicolumn{1}{c|}{0.70} & \multicolumn{1}{c|}{0.61} & \multicolumn{1}{c|}{0.65} \\
\multicolumn{1}{|l|}{\textemdash w/o Ensemble} & \multicolumn{1}{c|}{0.68} & \multicolumn{1}{c|}{0.66} & \multicolumn{1}{c|}{0.67} \\
\multicolumn{1}{|l|}{\textemdash w/o Question classifier} & \multicolumn{1}{c|}{0.61} & \multicolumn{1}{c|}{0.63} & \multicolumn{1}{c|}{0.62} \\\hline

\end{tabular}}

\end{center}
\end{table}

\subsection{RQ3: Supervision vs.~Indirect Supervision}

\begin{figure}[h]
    \centering
    \includegraphics[width=1\linewidth]{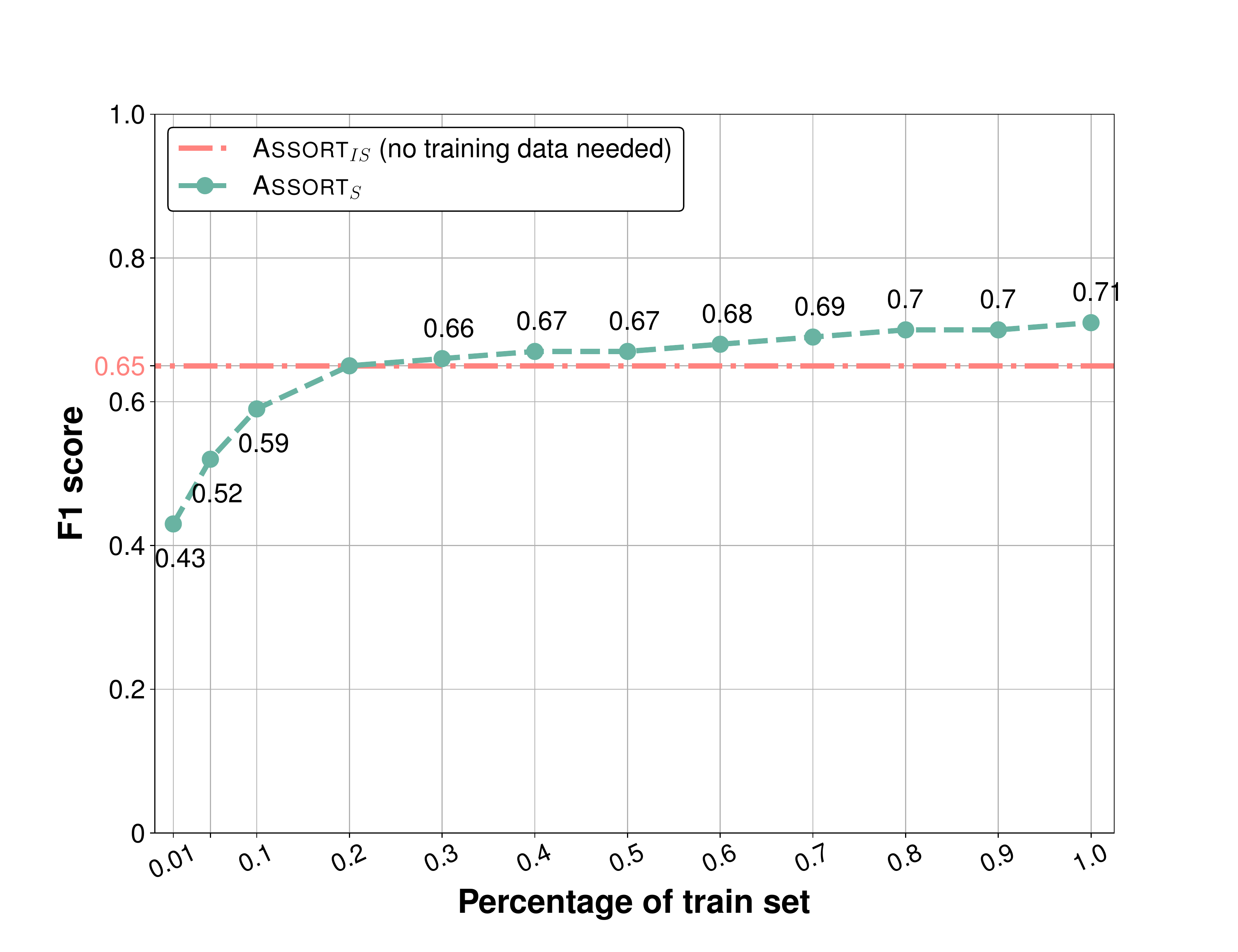}
    \caption{The F1 score of the direct vs.~indirect supervision methods when various amounts of training data are available.}
    \label{fig:supervised_vs_indirect_supervision}
\end{figure}

Figure~\ref{fig:supervised_vs_indirect_supervision} compares the accuracy of {\sname} and {\uname} when various amounts of training data are available. Given that {\uname} only uses pre-trained models and does not require any training data, the accuracy of {\uname} is constant (65\%) in these settings. When all training data (i.e., 2,424 posts and their summaries) is available, \sname{} outperforms \uname{} by 6\%. This indicates that directly training a supervised model is a better choice when there are sufficient training data. However, when 20\% or less of the original training data is available, {\uname} outperforms \sname{}. Therefore, when the training data is small, indirect supervision can be a better option than directly training a model with small training data. 
Furthermore, with an F1 score of 65\%,  \uname{}  outperforms all six comparison baselines by 7\% to 59\%. This result is significant since the best comparison baseline is fine-tuned on all training data. This implies that indirect supervision can be a promising yet low-cost alternative compared with unsupervised approaches and model fine-tuning. 

\subsection{RQ4: Human Evaluation}
\label{ablation}
\begin{figure}[t]
    \centering
    \includegraphics[width=\linewidth]{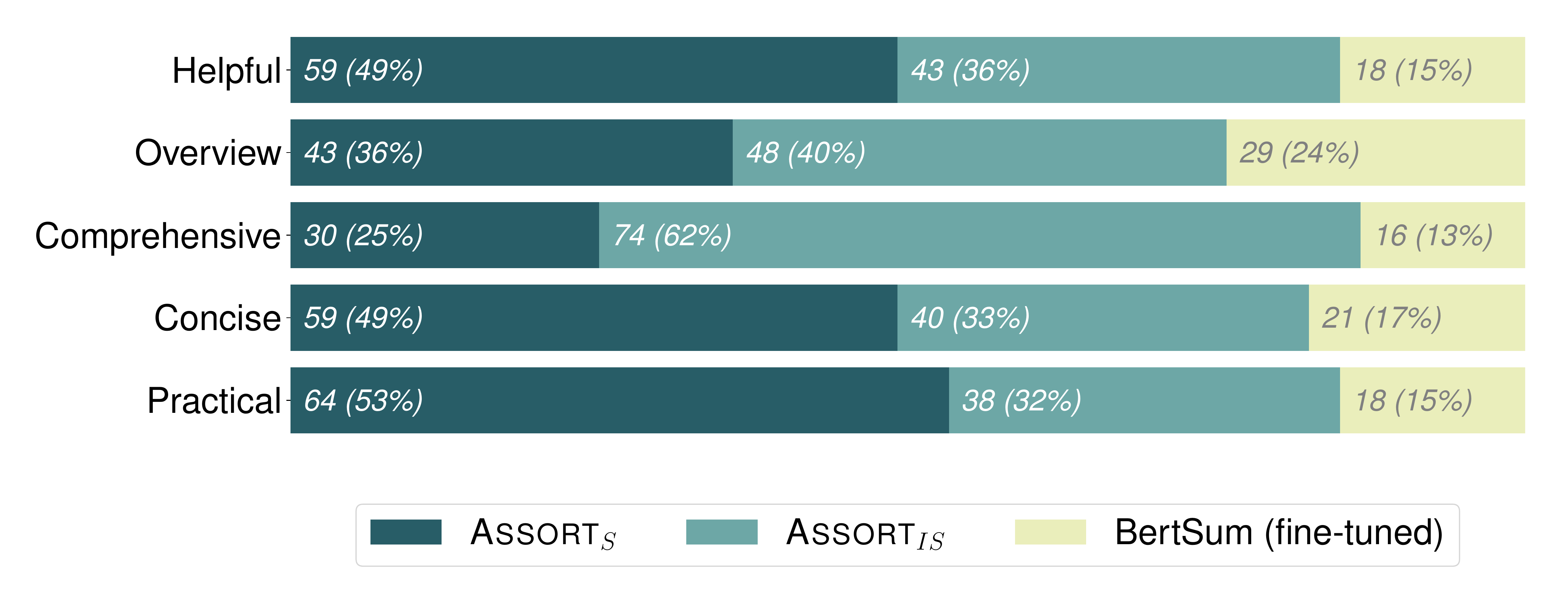}
    \caption{Participants' choices over different types of summaries in five aspects}
    \label{fig:bar_plot}
\end{figure}

Figure~\ref{fig:bar_plot} shows the choice of participants over the summaries generated by {\sname}, {\uname}, and {\em BertSum (fine-tuned)}  in five aspects. Overall, participants strongly preferred summaries generated by \sname{} or \uname{} over {\em BertSum (fine-tuned)}. Between \sname{} and \uname{}, more participants found summaries generated by \sname{} more helpful, concise, and practical. Yet more participants found summaries generated by \uname{} more comprehensive and providing a better overview of the post. 

Since each answer post and its summaries were reviewed by three participants, we further analyzed the consistency among those participants when they answered each multiple-choice question. In 60\% of the cases, the three participants assigned to the same post consistently chose summaries generated by \name{} over {\em BertSum (fine-tuned)} in a multiple-choice question. In 91\% of cases, at least two out of three participants consistently chose either \sname{} or \uname{} over {\em BertSum (fine-tuned)}. However, the choices between \sname{} and \uname{} were not very consistent among participants assigned to the same post. In only 40\% of cases, all three participants consistently chose \sname{} or \uname{}. This implies the participants chose between \sname{} and \uname{} largely based on their personal preference. Neither \sname{} or \uname{} really triumph over each other.

\begin{figure}[]
    \centering
    \includegraphics[width=.95\linewidth]{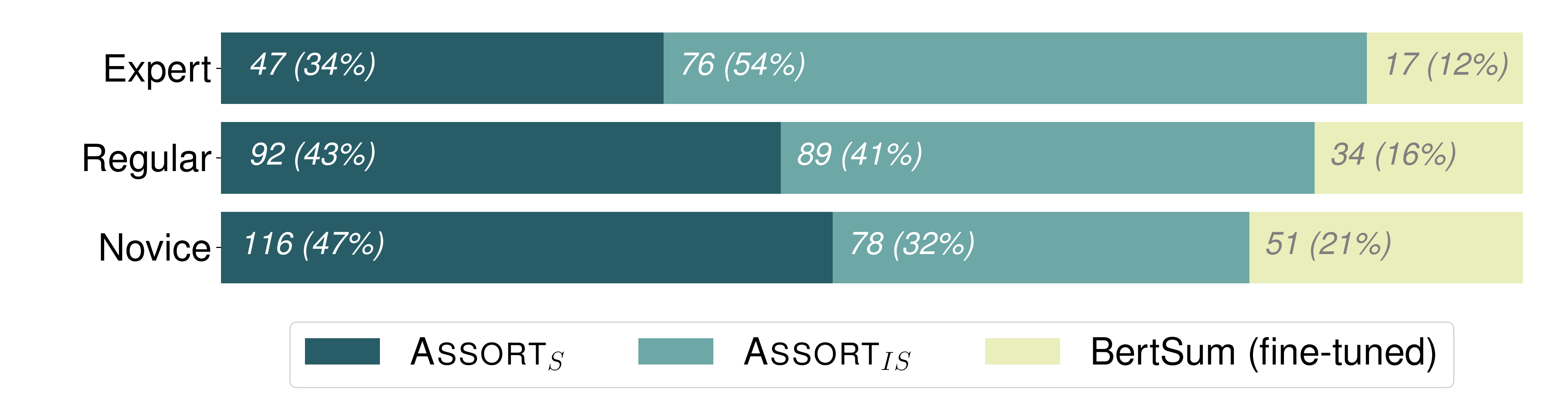}
    \caption{Choices of participants with different levels of expertise}
    \label{fig:bar_plot2}
\end{figure}

We also investigated how participants' expertise on concepts in a SO question affects their preferences over post summaries. As described in Section~\ref{sec:design}, participants first reported their expertise on a 7-point scale. We categorized their expertise as ``novice'' if their rating is 1-2, ``regular'' if their rating is 3-5, and ``expert'' if their rating is 6-7. Figure~\ref{fig:bar_plot2} shows the choices of participants with different levels of expertise. We observed an tendency of favoring \sname{} among novices while an tendency of favoring \uname{} among experts. One plausible reason for this is that novices prefer to see short summaries and feel overwhelmed if a summary contains too much information, while experts prefer to see more comprehensive information and feel less overwhelmed. Pearson's Chi-square test of independence shows that the choice difference among participants with different levels of expertise for post summaries is statistically significant ($p=0.0006$).

In the final survey, 84\% of participants confirmed they would like to see post summaries when browsing SO posts. For example, P4 elaborates, ``\emph{seeing a concise yet informative summary can help me quickly decide whether a post is worth reading or not.}'' Another participant (P10) said, ``\emph{most information in a long post is useless to me, what I want is the solution and solution only}''.
\section{discussion}
Our experiment shows that only fine-tuning a general text summarization model such as~\cite{liu2019fine} is insufficient. Our supervised approach, \sname{}, achieves significantly higher summarization accuracy by incorporating specific designs for SO posts, such as question classification and domain-specific features. This may carry a more general implication---as we reuse models from NLP or ML, we should consider renovating their model architectures and adapting them to account for the unique characteristics of SE tasks. 

Our indirect supervision approach, \uname{}, is proposed to address the challenge of lacking labeled data in SO post summarization. \uname{} has achieved reasonable accuracy and is proven to be a better choice when labeled data is insufficient, e.g., less than 485 posts in our experiment setting. Furthermore, our user study shows that while summaries generated by \uname{} and \sname{} were deemed good in different aspects (e.g., {\em comprehensiveness} vs.~{\em conciseness}), there was no significant preference difference in general. This implies that \uname{} could be an acceptable yet low-cost solution in practice. 

Both \sname{} and \uname{} can be generalized to summarize other types of SE documents, such as API documentation, tutorials, and bug reports. Since \uname{} only uses pre-trained models, it can be reused as-is. By contrast, \sname{} would benefit from several extensions, e.g., re-training a topic classifier rather than a question classifier, re-designing some domain-specific features based on the linguistic norms in the target corpus. Furthermore, since \uname{} does not require any labeled training data, it can also be used as a starting point to explore the feasibility and opportunities of summarizing other types of SE documents.

\textbf{\emph{Threats to validity.}} In terms of internal validity, labeling summative sentences in SO posts is a subjective task. Therefore, different labelers will not necessarily select the same set of summative sentences from the same post. In this work, two labelers expanded \dataset{} with 785 answer posts retrieved under 365 SO questions. We strictly follow the labeling procedure established in~\cite{kou2022sosum}. The Cohen’s kappa score is 0.72 on the initial labeling results before discussion, which indicates moderate agreement. Most disagreements were resolved after discussion. In the final dataset, only those sentences that both labelers agreed on are labeled as summative sentences.

In terms of external validity, since our dataset is constructed by retrieving posts under the most popular questions, it inevitably favors popular tags, such as Java and Python. Our dataset contains 1,089 unique tags, while there are more than 50K tags on Stack Overflow. The accuracy of \sname{} may drop a bit with unfamiliar topics due to unseen terminologies.  {Furthermore, our current model is only trained with SO posts to three common types of questions, since we aim to develop a proof of concept and demonstrate its feasibility for the scope of this work. To support other types of questions, one should enrich our dataset and re-train the model to obtain optimal accuracy.} Finally, the accuracy of \sname{} may also decrease when reused {\em as-is} for other types of SE documents. This is because \sname{} includes unique designs for SO posts, which is not applicable to other types of SE documents. 

In terms of construct validity, our question classifier does not always make the correct prediction. Sometimes, the boundary between different question categories can be blurred with questions like \emph{``How to fix headers-already-sent error in PHP?''}. This question should be classified as \textit{bug-fixing}, since it mentions an error. However, it is classified as \textit{how-to} by our question classifier since the question title contains a phrase \emph{``how to''}. To mitigate this threat, \sname{} includes an ensemble mechanism that merges predictions from separate summative sentence prediction models.  {Another threat to construct validity lies in the user study design. In the current design, participants were only asked to evaluate the quality of the post summaries with regard to the SO question. However, the usefulness of these summaries when being used to solve real programming tasks remains to be determined.}

\textbf{\emph{Future work.}} Currently, \sname{} only uses a  feedforward neural network as the classification head on top of BERT. Prior work~\cite{liu2019fine} has shown that the choice of classification heads has an influence on model accuracy. It would be worthwhile to further experiment with other types of classification heads, such as transformers and LSTMs. Furthermore, we have only considered single-document summarization in this work. It would be interesting to investigate how to perform multi-document summarization on a thread of SO posts. The challenge of multi-document summarization is not only to shorten the text, but also to organize information around the key aspects to represent diverse views. Compared with extractive summarization, abstractive summarization may be a more preferred solution, since it is capable of rephrasing sentences from different posts and generating gradual transitions. Diversity-based ranking algorithms such as Maximal Marginal Relevance (MMR)~\cite{carbonell1998use} can also be considered here. 


\section{Related Work}
\noindent\textbf{\emph{SO post summarization.}} 
Several approaches have been presented to summarize SO posts to concise texts to facilitate SO post navigation~\cite{answerbot, silva, wang2021automatic, essential_sentences}. AnswerBot extracts summative paragraphs from SO posts based on features such as information entropy and paragraph position~\cite{answerbot}. CraSolver uses a similar multi-factor ranking mechanism to summarize bug solutions~\cite{wang2021automatic}. Both AnswerBot and CraSolver generate summaries at the paragraph level. By contrast, our approach generates more fine-grained sentence-level summaries of SO posts. Nadi and Treude experimented with four different IR-based approaches to select essential sentences from SO posts~\cite{essential_sentences}. They conducted a survey with 43 developers and found that while participants would like to get navigation support for browsing SO, the IR-based approaches failed to provide such support. Motivated by these findings, we propose a novel DL-based framework that can more accurately identify summative sentences in SO posts in this work.  {Several approaches have been proposed to generate question titles for SO threads based on code snippets~\cite{gao2020generating, gao2021code2que}. However, since question title generation aims to summarize a question in a one-liner, these approaches cannot be applied to SO post summarization, which has a very different problem setting.}

\noindent\textbf{\emph{Text summarization for bug reports.}} There are several summarization techniques for other types of SE documents such as bug reports~\cite{rastkar2010summarizing, 10.1145/3387904.3389272, 6704866, mani2012ausum, li2018unsupervised}. 
Liu et al.~designed a new metric called \emph{believability score}, which measures the degree to which a sentence is
supported or refuted by other comments. Based on the idea of \emph{believability score}, they further developed BugSum, which selects summative sentences in a bug report~\cite{10.1145/3387904.3389272}. Rastkar et al.~created a small dataset with human-annotated summaries for 36 bug reports and used it to train a Logistic Regression classifier with 24 explicit features to identify summative sentences in a bug report~\cite{6704866}. 
Mani et al.~experimented with four unsupervised models--- Centroid, MMR, Grasshopper, 
and Diverse Rank---on the task of bug report summarization~\cite{mani2012ausum}. Unlike these techniques, we proposed two learning paradigms for building DL models for SO post summarization. 

\noindent\textbf{\emph{General text summarization.}} Many neural-based text summarization techniques have been proposed recently~\cite{liu2019fine, narayan, narayan2018ranking, dong2018banditsum, verma2017extractive, miller2019leveraging}. For example, Liu et al.~presented a BERT-based model and trained it on CNN/DailyMails~\cite{liu2019fine}. Narayan et al.~presented a CNN-based model that summarizes a single news article into a one-liner and trained it with XSum~\cite{narayan}. In another work, Narayan et al.~framed the task of text summarization as a ranking problem and proposed a global optimization method for training~\cite{narayan2018ranking}. Dong et al.~framed text summarization as a contextual bandit problem, in which each document is considered as a context and each combination of selected sentences is an action to take~\cite{dong2018banditsum}. While these models perform well on news articles, reusing these models in another domain, such as SO posts, is not an easy task due to domain shift. In this work, we propose \sname{}, a supervised model that accounts for unique characteristics of SO posts and demonstrate that it significantly outperforms a fine-tuned version of \liu{} (Section~\ref{sec:accuracy}).

\noindent\textbf{\emph{Other tool support for information seeking in SO.}} In addition to text summarization, many other kinds of tool support have been proposed to facilitate information seeking in SO~\cite{ponzanelli2013seahawk, ponzanelli2014mining, ye2014interrogative, zou2015learning, soliman2018improving, qr3, api, treude2016augmenting, li2018improving, subramanian2014live, opiner, pome, zhang2018code, reinhardt2018augmenting, zhou2016api}. For example, SeaHawk~\cite{ponzanelli2013seahawk} is an Eclipse plugin that integrates SO into an IDE. 
Chatbot4QR~\cite{9165927} expands a user query with tags of similar SO questions in order to improve search results. 
Ye et al.~\cite{ye2014interrogative} propose a re-ranking mechanism for SO search results of a user-defined query based on different search focus users may have.
They identify linguistic patterns in different types of queries and train separate models based on query types to re-rank search results.
SISE uses linguistic patterns to identify sentences that mention an API in SO posts and augments API documents with those sentences~\cite{treude2016augmenting}.
\section{conclusion}
In this work, we propose two complementary learning methods for automated post summarization in Stack Overflow (SO). {\sname} is a supervised method that accounts for unique characteristics in SO posts via question classification, domain-specific features, and ensemble inference. \uname{}, on the other hand, is an indirectly supervised method that does not require any labeled training data by leveraging a pre-trained model from another domain while addressing the domain shift issue via natural language inference. Both \sname{} and \uname{} significantly outperform six existing techniques in terms of summarization accuracy. \uname{} is demonstrated to be a promising yet low-cost solution in a low-resource setting. Furthermore, a user study shows that participants consistently favored summaries generated by \sname{} or \uname{} over the best baseline model, while the preference difference between \sname{} and \uname{} was small. In the future, we plan to extend \name{} to 
support multi-document summarization on the entire discussion thread
and also apply \name{} to other types of SE documents.

\section{data availability}
Our code and data have been made available in a GitHub repository.\footnote{https://github.com/BonanKou/ASSORT-Automatic-Summarization-of-Stack-Overflow-Posts}

\bibliographystyle{IEEEtran}
\bibliography{references}

\begin{thebibliography}{10}
\providecommand{\url}[1]{#1}
\csname url@samestyle\endcsname
\providecommand{\newblock}{\relax}
\providecommand{\bibinfo}[2]{#2}
\providecommand{\BIBentrySTDinterwordspacing}{\spaceskip=0pt\relax}
\providecommand{\BIBentryALTinterwordstretchfactor}{4}
\providecommand{\BIBentryALTinterwordspacing}{\spaceskip=\fontdimen2\font plus
\BIBentryALTinterwordstretchfactor\fontdimen3\font minus
  \fontdimen4\font\relax}
\providecommand{\BIBforeignlanguage}[2]{{%
\expandafter\ifx\csname l@#1\endcsname\relax
\typeout{** WARNING: IEEEtran.bst: No hyphenation pattern has been}%
\typeout{** loaded for the language `#1'. Using the pattern for}%
\typeout{** the default language instead.}%
\else
\language=\csname l@#1\endcsname
\fi
#2}}
\providecommand{\BIBdecl}{\relax}
\BIBdecl

\bibitem{interleaving_web}
J.~Brandt, P.~J. Guo \emph{et~al.}, ``Two studies of opportunistic programming:
  interleaving web foraging, learning, and writing code,'' in \emph{Proceedings
  of the SIGCHI Conference on Human Factors in Computing Systems}, 2009, pp.
  1589--1598.

\bibitem{how_programmers_search}
X.~Xia, L.~Bao \emph{et~al.}, ``What do developers search for on the web?''
  \emph{Empirical Software Engineering}, vol.~22, no.~6, pp. 3149--3185, 2017.

\bibitem{how_use_so_code}
Y.~Wu, S.~Wang \emph{et~al.}, ``How do developers utilize source code from
  stack overflow?'' \emph{Empirical Software Engineering}, vol.~24, no.~2, pp.
  637--673, 2019.

\bibitem{abdalkareem2017developers}
R.~Abdalkareem, E.~Shihab \emph{et~al.}, ``What do developers use the crowd
  for? a study using stack overflow,'' \emph{IEEE Software}, vol.~34, no.~2,
  pp. 53--60, 2017.

\bibitem{answerbot}
B.~Xu, Z.~Xing \emph{et~al.}, ``Answerbot: Automated generation of answer
  summary to developers' technical questions,'' in \emph{2017 32nd IEEE/ACM
  International Conference on Automated Software Engineering (ASE)}.\hskip 1em
  plus 0.5em minus 0.4em\relax IEEE, 2017, pp. 706--716.

\bibitem{essential_sentences}
S.~Nadi and C.~Treude, ``Essential sentences for navigating stack overflow
  answers,'' in \emph{2020 IEEE 27th International Conference on Software
  Analysis, Evolution and Reengineering (SANER)}.\hskip 1em plus 0.5em minus
  0.4em\relax IEEE, 2020, pp. 229--239.

\bibitem{silva}
R.~F. Silva, C.~K. Roy \emph{et~al.}, ``Recommending comprehensive solutions
  for programming tasks by mining crowd knowledge,'' in \emph{2019 IEEE/ACM
  27th International Conference on Program Comprehension (ICPC)}.\hskip 1em
  plus 0.5em minus 0.4em\relax IEEE, 2019, pp. 358--368.

\bibitem{huang2018api}
Q.~Huang, X.~Xia \emph{et~al.}, ``Api method recommendation without worrying
  about the task-api knowledge gap,'' in \emph{2018 33rd IEEE/ACM International
  Conference on Automated Software Engineering (ASE)}.\hskip 1em plus 0.5em
  minus 0.4em\relax IEEE, 2018, pp. 293--304.

\bibitem{qa}
D.~Khashabi, S.~Min \emph{et~al.}, ``Unifiedqa: Crossing format boundaries with
  a single qa system,'' \emph{EMNLP - findings}, 2020.

\bibitem{bart}
M.~Lewis, Y.~Liu \emph{et~al.}, ``Bart: Denoising sequence-to-sequence
  pre-training for natural language generation, translation, and
  comprehension,'' \emph{arXiv preprint arXiv:1910.13461}, 2019.

\bibitem{narayan2018ranking}
S.~Narayan, S.~B. Cohen \emph{et~al.}, ``Ranking sentences for extractive
  summarization with reinforcement learning,'' \emph{arXiv preprint
  arXiv:1802.08636}, 2018.

\bibitem{dong2018banditsum}
Y.~Dong, Y.~Shen \emph{et~al.}, ``Banditsum: Extractive summarization as a
  contextual bandit,'' \emph{arXiv preprint arXiv:1809.09672}, 2018.

\bibitem{liu2019text}
Y.~Liu and M.~Lapata, ``Text summarization with pretrained encoders,''
  \emph{arXiv preprint arXiv:1908.08345}, 2019.

\bibitem{dailymail}
``Document summarization on cnn/daily mail,''
  \url{https://paperswithcode.com/sota/document-summarization-on-cnn-daily-mail},
  2022, accessed: 2022-3-29.

\bibitem{kou2022sosum}
B.~Kou, Y.~Di, M.~Chen, and T.~Zhang, ``Sosum: A dataset of stack overflow post
  summaries,'' in \emph{2022 IEEE/ACM 19th International Conference on Mining
  Software Repositories (MSR)}.\hskip 1em plus 0.5em minus 0.4em\relax IEEE,
  2022, pp. 247--251.

\bibitem{liu2019fine}
Y.~Liu, ``Fine-tune bert for extractive summarization,'' \emph{arXiv preprint
  arXiv:1903.10318}, 2019.

\bibitem{robillard2015recommending}
M.~P. Robillard and Y.~B. Chhetri, ``Recommending reference api
  documentation,'' \emph{Empirical Software Engineering}, vol.~20, no.~6, pp.
  1558--1586, 2015.

\bibitem{DBLP:journals/corr/abs-1109-2128}
\BIBentryALTinterwordspacing
G.~Erkan and D.~R. Radev, ``Lexrank: Graph-based lexical centrality as salience
  in text summarization,'' \emph{CoRR}, vol. abs/1109.2128, 2011. [Online].
  Available: \url{http://arxiv.org/abs/1109.2128}
\BIBentrySTDinterwordspacing

\bibitem{brandt2009two}
J.~Brandt, P.~J. Guo, J.~Lewenstein, M.~Dontcheva, and S.~R. Klemmer, ``Two
  studies of opportunistic programming: interleaving web foraging, learning,
  and writing code,'' in \emph{Proceedings of the SIGCHI Conference on Human
  Factors in Computing Systems}, 2009, pp. 1589--1598.

\bibitem{xia2017developers}
X.~Xia, L.~Bao, D.~Lo, P.~S. Kochhar, A.~E. Hassan, and Z.~Xing, ``What do
  developers search for on the web?'' \emph{Empirical Software Engineering},
  vol.~22, no.~6, pp. 3149--3185, 2017.

\bibitem{gupta2019abstractive}
S.~Gupta and S.~K. Gupta, ``Abstractive summarization: An overview of the state
  of the art,'' \emph{Expert Systems with Applications}, vol. 121, pp. 49--65,
  2019.

\bibitem{huang2021factual}
Y.~Huang, X.~Feng, X.~Feng, and B.~Qin, ``The factual inconsistency problem in
  abstractive text summarization: A survey,'' \emph{arXiv preprint
  arXiv:2104.14839}, 2021.

\bibitem{kryscinski2019evaluating}
W.~Kry{\'s}ci{\'n}ski, B.~McCann, C.~Xiong, and R.~Socher, ``Evaluating the
  factual consistency of abstractive text summarization,'' \emph{arXiv preprint
  arXiv:1910.12840}, 2019.

\bibitem{cao2018faithful}
Z.~Cao, F.~Wei, W.~Li, and S.~Li, ``Faithful to the original: Fact aware neural
  abstractive summarization,'' in \emph{thirty-second AAAI conference on
  artificial intelligence}, 2018.

\bibitem{kryscinski2019neural}
W.~Kry{\'s}ci{\'n}ski, N.~S. Keskar, B.~McCann, C.~Xiong, and R.~Socher,
  ``Neural text summarization: A critical evaluation,'' \emph{arXiv preprint
  arXiv:1908.08960}, 2019.

\bibitem{goodrich2019assessing}
B.~Goodrich, V.~Rao, P.~J. Liu, and M.~Saleh, ``Assessing the factual accuracy
  of generated text,'' in \emph{proceedings of the 25th ACM SIGKDD
  international conference on knowledge discovery \& data mining}, 2019, pp.
  166--175.

\bibitem{falke2019ranking}
T.~Falke, L.~F. Ribeiro, P.~A. Utama, I.~Dagan, and I.~Gurevych, ``Ranking
  generated summaries by correctness: An interesting but challenging
  application for natural language inference,'' in \emph{Proceedings of the
  57th Annual Meeting of the Association for Computational Linguistics}, 2019,
  pp. 2214--2220.

\bibitem{treude2011programmers}
L.~B. De~Souza, E.~C. Campos \emph{et~al.}, ``Ranking crowd knowledge to assist
  software development,'' in \emph{Proceedings of the 22nd International
  Conference on Program Comprehension}, 2014, pp. 72--82.

\bibitem{question_type}
C.~Treude, O.~Barzilay \emph{et~al.}, ``How do programmers ask and answer
  questions on the web?(nier track),'' in \emph{Proceedings of the 33rd
  international conference on software engineering}, 2011, pp. 804--807.

\bibitem{rosen2016mobile}
C.~Rosen and E.~Shihab, ``What are mobile developers asking about? a large
  scale study using stack overflow,'' \emph{Empirical Software Engineering},
  vol.~21, no.~3, pp. 1192--1223, 2016.

\bibitem{allamanis2013and}
M.~Allamanis and C.~Sutton, ``Why, when, and what: analyzing stack overflow
  questions by topic, type, and code,'' in \emph{2013 10th Working conference
  on mining software repositories (MSR)}.\hskip 1em plus 0.5em minus
  0.4em\relax IEEE, 2013, pp. 53--56.

\bibitem{mchugh2012interrater}
M.~L. McHugh, ``Interrater reliability: the kappa statistic,'' \emph{Biochemia
  medica}, vol.~22, no.~3, pp. 276--282, 2012.

\bibitem{tabassum2020code}
\BIBentryALTinterwordspacing
J.~Tabassum, M.~Maddela, W.~Xu, and A.~Ritter, ``Code and named entity
  recognition in stackoverflow,'' in \emph{Proceedings of the 58th Annual
  Meeting of the Association for Computational Linguistics (ACL)}, 2020.
  [Online]. Available:
  \url{https://www.aclweb.org/anthology/2020.acl-main.443/}
\BIBentrySTDinterwordspacing

\bibitem{chen2017synonym}
C.~Chen, Z.~Xing, and W.~Ximing, ``Unsupervised software-specific morphological
  forms inference from informal discussions,'' in \emph{The 39th International
  Conference on Software Engineering, Buenos Aires, Argentina}.\hskip 1em plus
  0.5em minus 0.4em\relax IEEE, 2017.

\bibitem{api}
H.~Li, S.~Li \emph{et~al.}, ``Improving api caveats accessibility by mining api
  caveats knowledge graph,'' in \emph{2018 IEEE International Conference on
  Software Maintenance and Evolution (ICSME)}.\hskip 1em plus 0.5em minus
  0.4em\relax IEEE, 2018, pp. 183--193.

\bibitem{barthuggingface}
``Bart large cnn,'' \url{https://huggingface.co/facebook/bart-large-cnn}, 2022,
  accessed: 2022-1-7.

\bibitem{dagan2005pascal}
I.~Dagan, O.~Glickman, and B.~Magnini, ``The pascal recognising textual
  entailment challenge,'' in \emph{Machine learning challenges workshop}.\hskip
  1em plus 0.5em minus 0.4em\relax Springer, 2005, pp. 177--190.

\bibitem{yin2021docnli}
W.~Yin, D.~Radev, and C.~Xiong, ``Docnli: A large-scale dataset for
  document-level natural language inference,'' \emph{arXiv preprint
  arXiv:2106.09449}, 2021.

\bibitem{kingma2014adam}
D.~P. Kingma and J.~Ba, ``Adam: A method for stochastic optimization,''
  \emph{arXiv preprint arXiv:1412.6980}, 2014.

\bibitem{zhou2018neural}
Q.~Zhou, N.~Yang, F.~Wei, S.~Huang, M.~Zhou, and T.~Zhao, ``Neural document
  summarization by jointly learning to score and select sentences,'' in
  \emph{Proceedings of the 56th Annual Meeting of the Association for
  Computational Linguistics (Volume 1: Long Papers)}, 2018, pp. 654--663.

\bibitem{see2017get}
A.~See, P.~J. Liu, and C.~D. Manning, ``Get to the point: Summarization with
  pointer-generator networks,'' in \emph{Proceedings of the 55th Annual Meeting
  of the Association for Computational Linguistics (Volume 1: Long Papers)},
  2017, pp. 1073--1083.

\bibitem{celikyilmaz2018deep}
A.~Celikyilmaz, A.~Bosselut, X.~He, and Y.~Choi, ``Deep communicating agents
  for abstractive summarization,'' in \emph{Proceedings of the 2018 Conference
  of the North American Chapter of the Association for Computational
  Linguistics: Human Language Technologies, Volume 1 (Long Papers)}, 2018, pp.
  1662--1675.

\bibitem{sandhaus2008new}
E.~Sandhaus, ``The new york times annotated corpus,'' \emph{Linguistic Data
  Consortium, Philadelphia}, vol.~6, no.~12, p. e26752, 2008.

\bibitem{lexrank}
``Stack exchange data explorer,'' \url{https://pypi.org/project/lexrank/},
  2022, accessed: 2022-8-15.

\bibitem{wang2021automatic}
H.~Wang, X.~Xia \emph{et~al.}, ``Automatic solution summarization for crash
  bugs,'' in \emph{2021 IEEE/ACM 43rd International Conference on Software
  Engineering (ICSE)}.\hskip 1em plus 0.5em minus 0.4em\relax IEEE, 2021, pp.
  1286--1297.

\bibitem{carbonell1998use}
J.~Carbonell and J.~Goldstein, ``The use of mmr, diversity-based reranking for
  reordering documents and producing summaries,'' in \emph{Proceedings of the
  21st annual international ACM SIGIR conference on Research and development in
  information retrieval}, 1998, pp. 335--336.

\bibitem{gao2020generating}
Z.~Gao, X.~Xia \emph{et~al.}, ``Generating question titles for stack overflow
  from mined code snippets,'' \emph{ACM Transactions on Software Engineering
  and Methodology (TOSEM)}, vol.~29, no.~4, pp. 1--37, 2020.

\bibitem{gao2021code2que}
Z.~Gao, X.~Xia, D.~Lo, J.~Grundy, and Y.-F. Li, ``Code2que: A tool for
  improving question titles from mined code snippets in stack overflow,'' in
  \emph{Proceedings of the 29th ACM Joint Meeting on European Software
  Engineering Conference and Symposium on the Foundations of Software
  Engineering}, 2021, pp. 1525--1529.

\bibitem{rastkar2010summarizing}
S.~Rastkar, G.~C. Murphy, and G.~Murray, ``Summarizing software artifacts: a
  case study of bug reports,'' in \emph{2010 ACM/IEEE 32nd International
  Conference on Software Engineering}, vol.~1.\hskip 1em plus 0.5em minus
  0.4em\relax IEEE, 2010, pp. 505--514.

\bibitem{10.1145/3387904.3389272}
\BIBentryALTinterwordspacing
H.~Liu, Y.~Yu \emph{et~al.}, \emph{BugSum: Deep Context Understanding for Bug
  Report Summarization}.\hskip 1em plus 0.5em minus 0.4em\relax New York, NY,
  USA: Association for Computing Machinery, 2020, p. 94–105. [Online].
  Available: \url{https://doi.org/10.1145/3387904.3389272}
\BIBentrySTDinterwordspacing

\bibitem{6704866}
S.~Rastkar, G.~C. Murphy \emph{et~al.}, ``Automatic summarization of bug
  reports,'' \emph{IEEE Transactions on Software Engineering}, vol.~40, no.~4,
  pp. 366--380, 2014.

\bibitem{mani2012ausum}
S.~Mani, R.~Catherine, V.~S. Sinha, and A.~Dubey, ``Ausum: approach for
  unsupervised bug report summarization,'' in \emph{Proceedings of the ACM
  SIGSOFT 20th International Symposium on the Foundations of Software
  Engineering}, 2012, pp. 1--11.

\bibitem{li2018unsupervised}
X.~Li, H.~Jiang, D.~Liu, Z.~Ren, and G.~Li, ``Unsupervised deep bug report
  summarization,'' in \emph{2018 IEEE/ACM 26th International Conference on
  Program Comprehension (ICPC)}.\hskip 1em plus 0.5em minus 0.4em\relax IEEE,
  2018, pp. 144--14\,411.

\bibitem{narayan}
S.~Narayan, S.~B. Cohen \emph{et~al.}, ``Don’t give me the details, just the
  summary!'' \emph{Topic-aware Convolutional Neural Networks for Extreme
  Summarization. In}, 2018.

\bibitem{verma2017extractive}
S.~Verma and V.~Nidhi, ``Extractive summarization using deep learning,''
  \emph{arXiv preprint arXiv:1708.04439}, 2017.

\bibitem{miller2019leveraging}
D.~Miller, ``Leveraging bert for extractive text summarization on lectures,''
  \emph{arXiv preprint arXiv:1906.04165}, 2019.

\bibitem{ponzanelli2013seahawk}
L.~Ponzanelli, A.~Bacchelli, and M.~Lanza, ``Seahawk: Stack overflow in the
  ide,'' in \emph{2013 35th International Conference on Software Engineering
  (ICSE)}.\hskip 1em plus 0.5em minus 0.4em\relax IEEE, 2013, pp. 1295--1298.

\bibitem{ponzanelli2014mining}
L.~Ponzanelli, G.~Bavota \emph{et~al.}, ``Mining stackoverflow to turn the ide
  into a self-confident programming prompter,'' in \emph{Proceedings of the
  11th Working Conference on Mining Software Repositories}, 2014, pp. 102--111.

\bibitem{ye2014interrogative}
T.~Ye, B.~Xie, Y.~Zou, and X.~Chen, ``Interrogative-guided re-ranking for
  question-oriented software text retrieval,'' in \emph{Proceedings of the 29th
  ACM/IEEE international conference on Automated software engineering}, 2014,
  pp. 115--120.

\bibitem{zou2015learning}
Y.~Zou, T.~Ye, Y.~Lu, J.~Mylopoulos, and L.~Zhang, ``Learning to rank for
  question-oriented software text retrieval (t),'' in \emph{2015 30th IEEE/ACM
  International Conference on Automated Software Engineering (ASE)}.\hskip 1em
  plus 0.5em minus 0.4em\relax IEEE, 2015, pp. 1--11.

\bibitem{soliman2018improving}
M.~Soliman, A.~R. Salama, M.~Galster, O.~Zimmermann, and M.~Riebisch,
  ``Improving the search for architecture knowledge in online developer
  communities,'' in \emph{2018 IEEE International Conference on Software
  Architecture (ICSA)}.\hskip 1em plus 0.5em minus 0.4em\relax IEEE, 2018, pp.
  186--18\,609.

\bibitem{qr3}
M.~M. Rahman and C.~Roy, ``Effective reformulation of query for code search
  using crowdsourced knowledge and extra-large data analytics,'' in \emph{2018
  IEEE International Conference on Software Maintenance and Evolution
  (ICSME)}.\hskip 1em plus 0.5em minus 0.4em\relax IEEE, 2018, pp. 473--484.

\bibitem{treude2016augmenting}
C.~Treude and M.~P. Robillard, ``Augmenting api documentation with insights
  from stack overflow,'' in \emph{2016 IEEE/ACM 38th International Conference
  on Software Engineering (ICSE)}.\hskip 1em plus 0.5em minus 0.4em\relax IEEE,
  2016, pp. 392--403.

\bibitem{li2018improving}
H.~Li, S.~Li, J.~Sun, Z.~Xing, X.~Peng, M.~Liu, and X.~Zhao, ``Improving api
  caveats accessibility by mining api caveats knowledge graph,'' in \emph{2018
  IEEE International Conference on Software Maintenance and Evolution
  (ICSME)}.\hskip 1em plus 0.5em minus 0.4em\relax IEEE, 2018, pp. 183--193.

\bibitem{subramanian2014live}
S.~Subramanian, L.~Inozemtseva \emph{et~al.}, ``Live api documentation,'' in
  \emph{Proceedings of the 36th International Conference on Software
  Engineering}, 2014, pp. 643--652.

\bibitem{opiner}
G.~Uddin and F.~Khomh, ``Opiner: an opinion search and summarization engine for
  apis,'' in \emph{2017 32nd IEEE/ACM International Conference on Automated
  Software Engineering (ASE)}.\hskip 1em plus 0.5em minus 0.4em\relax IEEE,
  2017, pp. 978--983.

\bibitem{pome}
B.~Lin, F.~Zampetti \emph{et~al.}, ``Pattern-based mining of opinions in q\&a
  websites,'' in \emph{2019 IEEE/ACM 41st International Conference on Software
  Engineering (ICSE)}.\hskip 1em plus 0.5em minus 0.4em\relax IEEE, 2019, pp.
  548--559.

\bibitem{zhang2018code}
T.~Zhang, G.~Upadhyaya, A.~Reinhardt, H.~Rajan, and M.~Kim, ``Are code examples
  on an online q\&a forum reliable?: a study of api misuse on stack overflow,''
  in \emph{2018 IEEE/ACM 40th International Conference on Software Engineering
  (ICSE)}.\hskip 1em plus 0.5em minus 0.4em\relax IEEE, 2018, pp. 886--896.

\bibitem{reinhardt2018augmenting}
A.~Reinhardt, T.~Zhang \emph{et~al.}, ``Augmenting stack overflow with api
  usage patterns mined from github,'' in \emph{Proceedings of the 2018 26th ACM
  Joint Meeting on European Software Engineering Conference and Symposium on
  the Foundations of Software Engineering}, 2018, pp. 880--883.

\bibitem{zhou2016api}
J.~Zhou and R.~J. Walker, ``Api deprecation: a retrospective analysis and
  detection method for code examples on the web,'' in \emph{Proceedings of the
  2016 24th ACM SIGSOFT International Symposium on Foundations of Software
  Engineering}, 2016, pp. 266--277.

\bibitem{9165927}
N.~Zhang, Q.~Huang, X.~Xia, Y.~Zou, D.~Lo, and Z.~Xing, ``Chatbot4qr:
  Interactive query refinement for technical question retrieval,'' \emph{IEEE
  Transactions on Software Engineering}, vol.~48, no.~4, pp. 1185--1211, 2022.

\end{thebibliography}
\end{document}